&pdflatex

\documentclass[twocolumn,showpacs]{revtex4}

\usepackage{graphicx}
\usepackage{amsmath}
\usepackage{amsfonts}
\usepackage{amssymb}

\begin{document}
	\title{Generation of families of spectra in $\mathcal{PT}$-symmetric quantum mechanics and scalar bosonic field theory}
	\author{Steffen Schmidt}
	\email{steffen@ppmss.de}
	\author{S. P. Klevansky}
	\email{spk@physik.uni-heidelberg.de}
	\affiliation{Institut f\"{u}r Theoretische Physik, Universit\"{a}t Heidelberg, Philosophenweg 19, 69120 Heidelberg, Germany}
	
	\begin{abstract}
		This paper explains the systematics of the generation of families of spectra for the $\mathcal{PT}$-symmetric quantum-mechanical Hamiltonians 
		$H=p^2+x^2(ix)^\epsilon$, $H=p^2+(x^2)^\delta$, and $H=p^2-(x^2)^\mu$. In addition, it contrasts the results obtained with those found for a bosonic
		scalar field theory, in particular in one dimension, highlighting the similarities and differences to the quantum-mechanical case. It is shown 
		that the number of families of spectra can be deduced from the number of noncontiguous pairs of Stokes' wedges that display 
		$\mathcal{PT}$-symmetry. To do so, simple arguments that use the WKB approximation are employed, and these imply that the eigenvalues are real.  However, definitive results are in most cases presently only obtainable numerically, and  not all eigenvalues in each family may be real.   Within the approximations used, it is illustrated that the difference between the quantum-mechanical and the field-theoretical cases lies in the number of accessible regions in which the eigenfunctions decay exponentially. This paper reviews and implements well-known techniques in complex analysis and $\mathcal{PT}$-symmetric quantum theory.  
	\end{abstract}
	\maketitle
	\section{Introduction}
	\label{Intro}
		Since the original work of Bender and Boettcher \cite{BB}, there have been many studies of $\mathcal{PT}$-symmetric systems and in the last few years 
		the literature on such systems has expanded rapidly. In particular, in the seminal paper mentioned above, the authors showed that a set or spectrum of real eigenvalues could be obtained for $\mathcal{PT}$-symmetric Hamiltonians of the form $H=p^2+x^2(ix)^\epsilon$, the "extended $\mathcal{PT}$ oscillator". They found that, when $\epsilon\ge0$, $\mathcal{PT}$-symmetry is realised as a manifest symmetry and consequently all eigenvalues are real; for $-1 < \epsilon< 0$, the $\mathcal{PT}$-symmetry is partially broken, and only a finite number of real eigenvalues exist. Below $-1$, the $\mathcal{PT}$-symmetry may be completely broken; however, the situation is this case has not yet been examined in detail.   The feature of a phase transition between a manifest and a broken phase is a natural consequence that can be expected in  any theory displaying two different phases. Such a phase transition has been a major focus of investigation and details can be found in \cite{BBM1999} for this set of Hamiltonians.  A further discussion can be found in Bender's review article, in which several  examples are presented \cite{B2007}. Several other papers published since then explore the properties of one-dimensional systems, such as the $\mathcal{PT}$-symmetric square-well problem \cite{BT2006}, or $\mathcal{PT}$-symmetric matrix Hamiltonians, see examples in \cite{B2007}. In all of these papers, the aim was to elucidate the point that $\mathcal{PT}$-symmetric Hamiltonians, even if they appear nonconventional, can display a spectrum of
		 real eigenvalues and that a phase transition from a symmetric to a broken phase can occur.

		Recently,  it was argued that bosonic field theories described by the (Euclidean) Lagrangian 
		\begin{equation}																													\mathcal{L}=\frac{1}{2}(\partial\phi(x))^2+\frac{g}{4n+2}\phi^{4n+2}(x)-J(x)\phi(x),
		\label{eq:lag} 
		\end{equation}
		where $n$ is a positive integer, can admit {\it several  families} of real eigenvalues \cite{BK2010}. This raises the 
		question as to whether the quantum-mechanical Hamiltonians $H$ for the extended  $\mathcal{PT}$-symmetric oscillator admit several real spectra and not just one for each value of $\epsilon$ as shown in \cite{BB}, and, if so, whether one can classify the number of families of solutions that are generated. In this paper, we address this question for the Hamiltonians $H_I=p^2+x^2(ix)^\epsilon$, 			$H_{II}=p^2+(x^2)^\delta$ \cite{BMPS1989} and $H_{III}=p^2-(x^2)^\mu$, with $\epsilon, \delta$ and $\mu >0, \in Z$, and thereafter discuss the 
systematics in the field-theoretical case for the Lagrangian in Eq.~(\ref{eq:lag}), for different values of $n$. In doing so, we also aim at giving a primer on the techniques of complex analysis used, so that this paper is self-contained and can be regarded as an introduction to this subject.

		This paper is constructed as follows. In Section \ref{Basic}, we give an overview of the results, review the basic methodology of analysing the asymptotic behaviour
		of the eigenfunctions in the complex plane, and discuss how families of real solutions can arise. The analysis is based on the WKB approximation \cite{WKB}. Exact results are presently only obtainable numerically and a rigorous proof that the eigenvalues are all  real is difficult. Analytic continuation of the eigenvalue problem has been discussed, for example, for the anharmonic oscillator in \cite{BT1993}.

		In Section \ref{mech}, we apply the methodology to the Hamiltonians $H_I$, $H_{II}$ and $H_{III}$ for positive integral values of $\epsilon$, $\delta$ and $\mu$, and in 
		Section \ref{field} to the generalized field-theoretic bosonic Lagrangian, with n taking values from 0 to 2 in steps of $1/4$.
	\section{Basic Methodology}
	\label{Basic}

		Symmetry in parity and time is expressed by the four relations $P\hat x P = -\hat x$, $P\hat p P = - \hat p$, $T \hat x T = \hat x$ and $T \hat p T= - \hat p$, and  $T$ 
		is an antilinear operator, $TiT= -i$. Using this prescription, it is easily verified that $H_I$, $H_{II}$ and $H_{III}$ are all $\mathcal{PT}$-symmetric. It is also obvious that the quantum-mechanical Hamiltonians $H_{II}$ and $H_{III}$ are Hermitian in the Dirac sense, i.e. $H=H^\dagger$. This is also true for $H_I$, for certain values of $\epsilon$. It is the case, for example, when $\epsilon=0,4,8,...$; then $H_{I,\epsilon=0}=p^2+x^2$; $H_{I, \epsilon=4}=p^2+x^6$; $H_{I,\epsilon=8}=p^2+x^{10}$, and so on. These Hamiltonians all appear bounded from below, and each is known to possess a spectrum of real eigenvalues. However, by choosing appropriate $\mathcal{PT}$-symmetric boundary conditions, one can show that these Hamiltonians (with $\epsilon =0, 4, 8, ...$) in fact have $N=(\epsilon+4)/4$ different sets of eigenvalues or families, which in the WKB approximation take on real values. That is, the first Hamiltonian, $H_{I,\epsilon=0}$, has only one real spectrum, $H_{I,\epsilon=4}$ may display two families of solutions, or have two real spectra, and $H_{I,\epsilon=8}$ may have three families with real solutions, or three possible spectra of real eigenvalues. 

		Continuing to examine $H_I$, one notes that for $\epsilon=2,6,10,...$, these operators take on the form $H_{I,\epsilon=2}=p^2-x^4$, $H_{I,\epsilon=6}= p^2-x^8$, $H_{I, \epsilon=10}= p^2-x^{12}$, and so on, all of which are Hermitian, but which otherwise appear unbounded from below. This set of Hamiltonians, however, can also exhibit spectra of real eigenvalues when $\mathcal{PT}$-symmetric boundary conditions are imposed. Here the number of families that can occur is given by $N=(\epsilon+2)/4$. That is,  $H_{I, \epsilon=2}$ may possess a single spectrum of real eigenvalues, $H_{I,\epsilon=6}$ may possess two real spectra, while $H_{I,\epsilon=10}$ may have three.

		Finally, the last possible choices of $\epsilon$ for $H_I$, $\epsilon$ odd, lead to Hamiltonians that are manifestly 
$\mathcal{PT}$-symmetric but not Hermitian: $H_{I,\epsilon=1}=p^2+ix^3$, $H_{I,\epsilon=3}=p^2-ix^5$, $H_{I,\epsilon=5}=p^2+ix^7$ \dots. In these cases $N=(\epsilon+1)/2$ families appear that can have real eigenvalues.		

		The basic methodology underlying these results and which is generally applicable to other systems involves an analysis of the eigenfunctions in the complex plane.
Both in $\mathcal{PT}$-symmetric quantum mechanics and in scalar field theory, integrals occur whose integration paths can, in the most general case, be taken to lie in the complex plane. In quantum mechanics, the wave function is obtained by integrating the Schr\"odinger equation and imposing boundary conditions, while in field theory, the properties of the system can be derived from the functional integral formulation of the action in terms of the Lagrangian. 
		In each case, one has to analyse the convergence properties of the integral; independent paths in the complex plane correspond to the possible existence of individual families with spectra containing real eigenvalues.
		In the following it is described how, with the help of the symmetries of the regions of convergence of the controlling asymptotic factors, the number of real spectra possible both in 
		quantum mechanics and in quantum field theory can be deduced. We follow the methodology sketched in \cite{B2007}.\\
		
\begin{figure}[ht]
			\centering
				\includegraphics[width=0.35\textwidth]{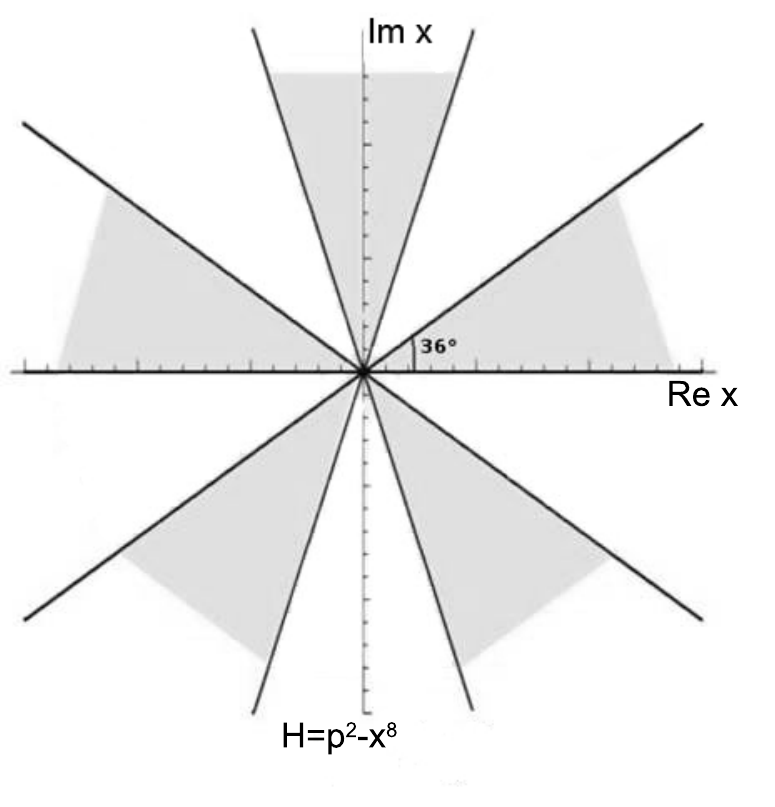}
			\caption{Regions of convergence for the asymptotic wave function associated with $H  _{I,\epsilon=6}=p^2-x^8$ (Stokes' wedges).}
			\label{fig:x80} 
		\end{figure}

		The first step in dealing with the problem is to identify the regions of convergence of the asymptotic wave function, known as Stokes' wedges.  Figure \ref{fig:x80} shows these regions associated with the Hamiltonian $H _{I,\epsilon=6}=p^2-x^8$, as an example.
	
  As there are always two linearly independent solutions of the Schr\"odinger equation, one finds that the exponential behaviour of  one solution decays in certain regimes (denoted as unshaded Stokes' wedges), while the second solution grows in this region. Alternately, the exponential part of the second solution decays in the shaded regions, while the first solution grows. Thus, a wave function can always be found whose exponential  behaviour converges to zero for large $|x|$ in some wedge of the complex plane, and the shaded and unshaded regions
		reflect the difference in sign occurring in the exponent that causes the falling or rising asymptotic behaviour. The lines separating the wedges are called Stokes' lines and these are the lines along which the solution is purely oscillatory. An integration path, however, must lie in regions in which the solution converges to zero asymptotically. Many paths are possible, but not all necessarily lead to a spectrum that can contain real eigenvalues.  We conjecture that {\it the number of paths that
		can possibly lead to distinct real solutions is given by the number of pairs of noncontiguous $\mathcal{PT}$-symmetric Stokes' wedges, i.e. Stokes' wedges that are 
		symmetric to the imaginary axis, while two pairs of Stokes' wedges which are symmetric to the real axis may only be counted once.} Hence 
		an integration path leads to a spectrum with real eigenvalues only if it ends in  Stokes' wedges that are $\mathcal{PT}$-symmetrically arranged. 

In the case shown in Fig.~\ref{fig:x80}, following the conjecture above, we see by inspection that the system can admit two different real spectra: the shaded pair of Stokes' wedges lying under the real axis are $\mathcal{PT}$-symmetric to one another. This is also the case for the two unshaded Stokes' wedges lying directly below the real axis. The reflection of this structure into the upper half-plane does not bring new solutions. As has already been noted, this case is special in that in Hermitian quantum mechanics it has no bound states and it only displays a continuous real spectrum, as it is not bounded below. As one can see, the real axis is not an admissible integration path, reflecting this behaviour. In this example, however, we note that there are 10 different wedges in the complex plane. In the next section we will examine the number of classical turning points - in this case there are 8. However, only two pairs can be identified that are  $\mathcal{PT}$-symmetric to one another, meaning that two possible spectra of real eigenvalues can exist. This forms the basis of the conjecture made above.

\begin{figure}[ht]
		\begin{center}
			\includegraphics[width=0.35\textwidth]{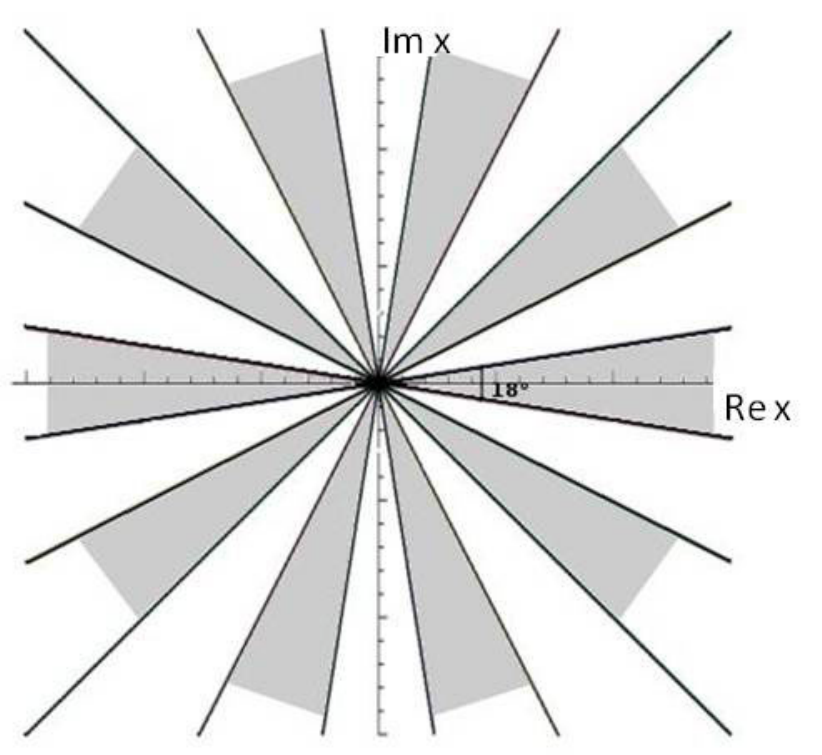}
			\caption{\label{fig:fi100}  Regions of convergence for the $n=2$ one dimensional scalar field theory (Stokes' wedges), shaded in grey.}
		\end{center}
		\end{figure}

In scalar quantum field theory the number of real spectra can be ascertained in a similar fashion. In apposition to quantum mechanics, where the sign of the determining integral comes from the wave function and its asymptotic behaviour always allows for a convergent solution, the sign in quantum field theory is fixed. This is due to the fact that the determining integral is the Euclidean action, which has a fixed sign. Thus, one finds alternating regions of convergence and divergence, limiting the number of families of solutions that a bosonic field theory can give rise to. 
Figure \ref{fig:fi100} shows the pattern of regions of convergence and divergence for the Lagrangian of Eq.~(\ref{eq:lag}) when n=2 in one dimension, i.e. for an 					interaction proportional to $\phi^{10}$.
	 In this case, the unshaded areas refer to divergent regions. The integration path may thus only lie in the shaded regions. As in quantum mechanics, only
		paths lying in pairs of $\mathcal{PT}$-symmetric Stokes' wedges lead to real eigenvalues. As a consequence, the number of real families of solutions is also 
		given by the number of unique pairs of $\mathcal{PT}$-symmetric Stokes' wedges. In this case, it is seen by inspection that there are three pairs of convergent wedges that are arranged
$\mathcal{PT}$-symmetrically. Thus three families or spectra can be identified.

	\section{Families of solutions in $\mathcal{PT}$-symmetric quantum mechanics}
	\label{mech}

		In this Section we apply the procedure outlined in Section \ref{Basic}  to the quantum-mechanical 
		Hamiltonians $H_I=p^2+x^2(ix)^\epsilon$, $H_{II}=p^2+(x^2)^\delta$ and $H_{III}=p^2-(x^2)^\mu$, $\epsilon, \delta, \mu>0, \in Z$ to determine the number of possible real spectra or families of real eigenvalues that can arise as a consequence of $\mathcal{PT}$-symmetry within the WKB approximation \cite{WKB}.  We review the general procedure of determining the 
		regions of convergence in the complex plane and the energy eigenvalues within this approximation.\\

		As indicated in the last section, our first task is to determine the regions in which the wave function converges in the complex $x$ plane, such as shown for example in Fig.~\ref{fig:x80} for $H_{I,\epsilon=6}$.
Determining the regions of convergence does not require a detailed solution of the associated Schr\"odinger equation - it is sufficient to investigate the asymptotic behaviour of the wavefunction as  $|x|\rightarrow \infty$, where $x$ is regarded as a complex variable \cite{B2007}. Within the 
		 WKB approximation, this limit  is determined by the exponential controlling factor of the asymptotic wavefunction, 
		\begin{equation}
			\psi(x)\sim\exp\left(\pm \int dx \sqrt{V(x)}\right).
		\label{eq:WKBWavefunction}
		\end{equation}
		 To analyse in which regions of the complex plane this function converges to zero when $|x|\rightarrow \infty$, one determines the lines on which the wavefunction is purely oscillatory, i.e. where its real part vanishes. These Stokes' lines are thus found by  evaluating Eq.~(\ref{eq:WKBWavefunction}) for the appropriate potential, setting $x=r\exp (i\varphi)$, with $|x|=r$, and requiring that the real part vanish on taking the limit $|x|\rightarrow\infty$. This condition leads to the Stokes' angle $\varphi_S$; the opening angle of each wedge is given by  $\theta=2\varphi_S$.

To  determine thereafter which solution decays or grows within each wedge, the anti-Stokes' lines are evaluated using Eq.~(\ref{eq:WKBWavefunction}) and imposing the condition that the imaginary part of the integral vanishes. This leads to a condition for $\varphi_A$. 

		Now making further use of the WKB approximation, one can determine the real (positive) energy eigenvalues associated with each potential, noting that
		\begin{equation}
			\int\limits_{x_{-}}^{x_{+}} dx \sqrt{E_n-V(x)}=(n+1/2)\pi,
		\label{eq:energy}
		\end{equation}
		where $x_{-}$ and $x_{+}$ are the classical turning points evaluated via the condition $E_n=V(x)$. If $x$ is complex, there can be several turning points. In the next subsection, we will explicitly show that if the complex pairs of turning points are chosen to be $\mathcal{PT}$-symmetric, the WKB approximation leads to the occurrence of real eigenvalues. The number of $\mathcal{PT}$-symmetric complex pairs of turning points thus determines the number of families containing real WKB eigenvalues.

\subsection{Hamiltonians of the form $H_I=p^2+x^2(ix)^\epsilon$.}	

In the following, we  give the results obtained for Hamiltonians of the form $H_I=p^2+x^2(ix)^\epsilon$. As $|x|\rightarrow\infty$, the controlling behaviour of the asymptotic wavefunction is given by
		\begin{equation}
		\psi\sim \exp{ \left[\mp\frac {2}{\epsilon+4} r^{\frac{\epsilon+4}{2}}e^{i(\frac{\pi}{2}+\phi)\frac{\epsilon+4}{2}}  \right]},
		\label{eq:psiHI}
		\end{equation}
so that the Stokes' lines are given by
		\begin{equation}
	           \varphi_S = \left(k+\frac{1}{2}\right)\frac{2}{\epsilon+4}\pi -\frac{\pi}{2}
		\label{eq:stokes}
		\end{equation}
and the anti-Stokes' lines by
		\begin{equation}
		\varphi_A=\frac{2k}{\epsilon+4}\pi-\frac{\pi}{2}.
		\label{eq:antistokes}
		\end{equation}
The opening angle of the wedges is given by
		\begin{equation}
		\theta= 2\varphi_S (k=0) = \frac{2}{\epsilon+4}\pi -\pi.
		\label{eq:openingangleHI}
		\end{equation}

The energy spectra associated with the potential $x^2(ix)^\epsilon$ are obtained by evaluating Eq.~(\ref{eq:energy}). To do this, it is first necessary to evaluate the turning points
$x_-$ and $x_+$. Equating $E_n$ with the potential (assuming real and positive energies), leads to the condition
  		\begin{equation}
			E_n=-(ix)^{\epsilon+2}, 
		\label{eq:ens}
		\end{equation}
which has $\epsilon+2$ possible solutions for  $x_\pm$. In general they can be written in the form
		\begin{equation}
			x_+= E_n^{1/(\epsilon+2)} e^{i\gamma}, \quad x_-=E_n^{1/(\epsilon+2)} e^{i\zeta},
		\label{eq:upperlimit}
		\end{equation}
where $\gamma$ and $\zeta$ are angles in which the asymptotic wave function converges and which are determined by the potential. Inserting this {\it ansatz} into
Eq.~(\ref{eq:energy}) and making suitable transformations in the integrals from $x_-$ to $0$ and from $0$ to $x_+$, one arrives at the result
		\begin{equation}
		(n+1/2)\pi = E_n^{(\epsilon+4)/(2\epsilon+4)}(e^{i\gamma} - e^{i\zeta})\int_0^1 dy\sqrt{1-y^{\epsilon+2}}.
		\label{eq:intermediatestep}
		\end{equation}

It follows from this relation that the spectrum can only be real if $\zeta=\pi-\gamma$, i.e. if 
                       \begin{equation}
 	           x_-=E_n^{1/(\epsilon+2)} e^{i(\pi-\gamma)}.     
		\label{eq:ll}
		\end{equation}
Thus the turning points $x_+$ and $x_-$ must lie     $\mathcal{PT}$-symmetrically to one another in order for the eigenvalues to be real. In addition, this result is independent of the sign of gamma, i.e. it is reflection-symmetric.   With this choice of $\zeta$, it follows that
		\begin{equation}
                        E_n^{(\epsilon+4)/(2\epsilon+4)}    2\cos (\gamma)\int_0^1   dy\sqrt{1-y^{\epsilon+2}}=(n+1/2)\pi.       
		\end{equation}
Evaluating the integral in the above expression finally leads to the closed form
		\begin{equation}
		E_n(\gamma) = \left[\frac{\Gamma\left(\frac{3}{2} + \frac{1}{\epsilon +2}\right)\sqrt\pi\left(n+\frac{1}{2}\right)}{\cos(\gamma)\Gamma\left(1+\frac{1}{\epsilon+2}\right)}\right]^{(2\epsilon+4)/(\epsilon+4)},
		\label{eq:engamma}
		\end{equation}
with the consequence that the relationship between two families of spectra arising from different wedges with convergence angles $\gamma$ and $\gamma^\prime$ is given by 
		\begin{equation}
		\frac{E_n(\gamma^\prime)}{E_n(\gamma)} = \left[\frac{\cos(\gamma)}{\cos(\gamma^\prime)}\right]^{(2\epsilon+4)/(\epsilon+4)}.
		\label{eq:ratioenergies}
		\end{equation}

Examining first the convergence behaviour in the complex plane, one finds that it differs for the cases in which $\epsilon$ is odd, even and divisible by four, or even and not divisible by four. These three cases are illustrated separately.

		\begin{widetext}

    		\begin{figure}[h]
			\begin{center}
			\includegraphics{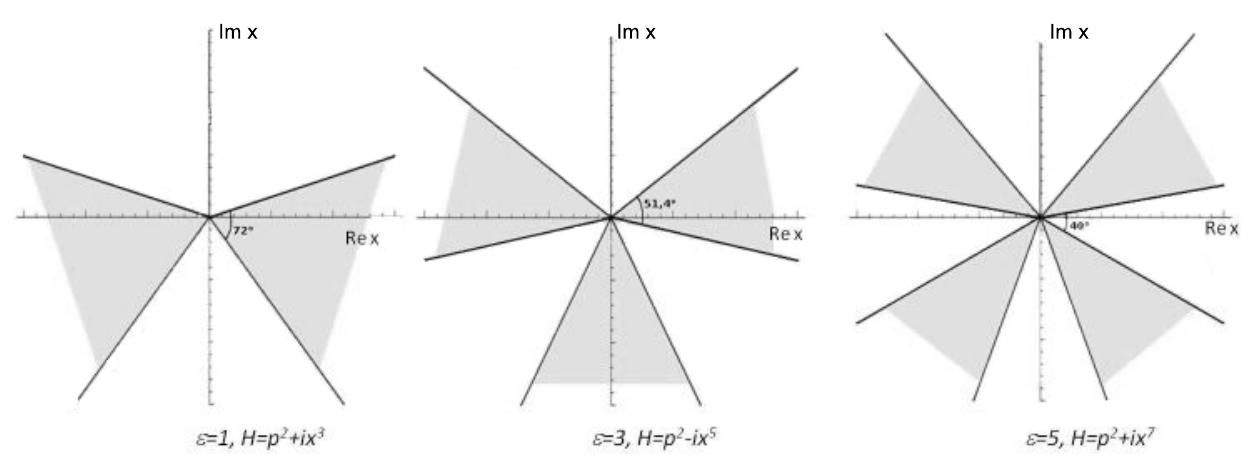}
			\end{center}
			\caption{Regions of convergence (Stokes' wedges) for odd values of $\epsilon = 1,3,5$.}
			\label{fig:PicQM1}  
		\end{figure}		
		\end{widetext}

The case in which $\epsilon$ is odd and the Hamiltonian is manifestly non-Hermitian is shown in Fig.~\ref{fig:PicQM1} for $\epsilon=1$, $3$, $5$. 
When $\epsilon=1$, and $V(x)=ix^3$, there are three possible turning points, $x_\pm = E_n^{1/3}\exp(-i\pi/6)$,  $E_n^{1/3}\exp(-i5\pi/6)$,  $E_n^{1/3}\exp(i\pi/2)$. Only one pair of these values lies $\mathcal{PT}$-symmetrically to one another. This pair, $x_-=E_n^{1/3}\exp(-i5\pi/6)$ and $x_+=E_n^{1/3}\exp(-i\pi/6)$,  lies within the shaded areas of the diagramme, and is thus the only pair that can lead to a real spectrum via the WKB calculation. Continuing in this fashion, when $\epsilon=3$ and $V(x)=-ix^5$, five possible turning points can be identified at angles $i\pi/10, i\pi/2, -3i\pi/10, -7i\pi/10, -11i\pi/10$. Now two pairs are $\mathcal{PT}$-symmetric to one another, the pair $x_-=E_n^{1/5}\exp(-11i\pi/10)$ and $x_+=E_n^{1/5}\exp(i\pi/10)$ and the pair
$x_-=E_n^{1/5}\exp(-7i\pi/10)$ and $x_+=E_n^{1/5}\exp(-3i\pi/10)$. The first pair of turning points lies in the shaded region of the diagramme in Fig.~\ref{fig:PicQM1}, while the second lies in the unshaded pair of wedges below the real axis. Each of these pairs of turning points lying in each of these pairs of wedges gives rise to a WKB spectrum of real eigenvalues. For $\epsilon=5$, this pattern continues, and seven turning points are identified; these can be ordered into three pairs that lie $\mathcal{PT}$-symmetrically to one another and thus lead to three distinct real spectra, according to the WKB prescription.

These results can be summarised into a heuristic prescription: by examining  Fig.~\ref{fig:PicQM1}, we note that
for all of these cases, one pair of wedges in which the controlling factor of the asymptotic solution converges to zero includes the real axis and is $\mathcal{PT}$-symmetric, corresponding to the generation of one real spectrum. This is the spectrum that was found and discussed in Refs.~\cite{BB} and \cite{B2007}. In the case when $\epsilon=3$, another possible real spectrum develops, which can be associated with the additional pair of $\mathcal{PT}$-symmetric wedges (unshaded) that has developed in the lower half-plane. For $\epsilon=5$, yet a further possible real spectrum develops, and this is associated with the next pair of noncontiguous $\mathcal{PT}$-symmetric wedges (shaded) that has developed in the upper half-plane. Thus we note that the number of spectra that can form and that is actually determined by the number of $\mathcal{PT}$-symmetric pairs of turning points can be inferred from the convergence pattern of Stokes' wedges in the complex plane. The general rule is that this number corresponds to the number of noncontinguous $\mathcal{PT}$-symmetric pairs of Stokes' wedges. We will simply make use of this observation in the next sections.

To sum up the specific results for odd values of $\epsilon$, we find that the general number of real spectra that can be generated in this case is given as $N=(\epsilon+1)/2$. Using Eq. ~(\ref{eq:engamma}) we find for $\epsilon=1$, $E_n\propto (n+1/2)^{6/5}$; for $\epsilon=3$, $E_n\propto (n+1/2)^{10/7}$ and the two spectra have the ratio $R=1.99:1$. For $\epsilon=5$, $E_n\propto (n+1/2)^{14/9}$ and the ratio of the resulting spectra is given as $R= 3.52:1.41:1$.

		\begin{widetext}

    		\begin{figure}[h]

			\begin{center}
			\includegraphics{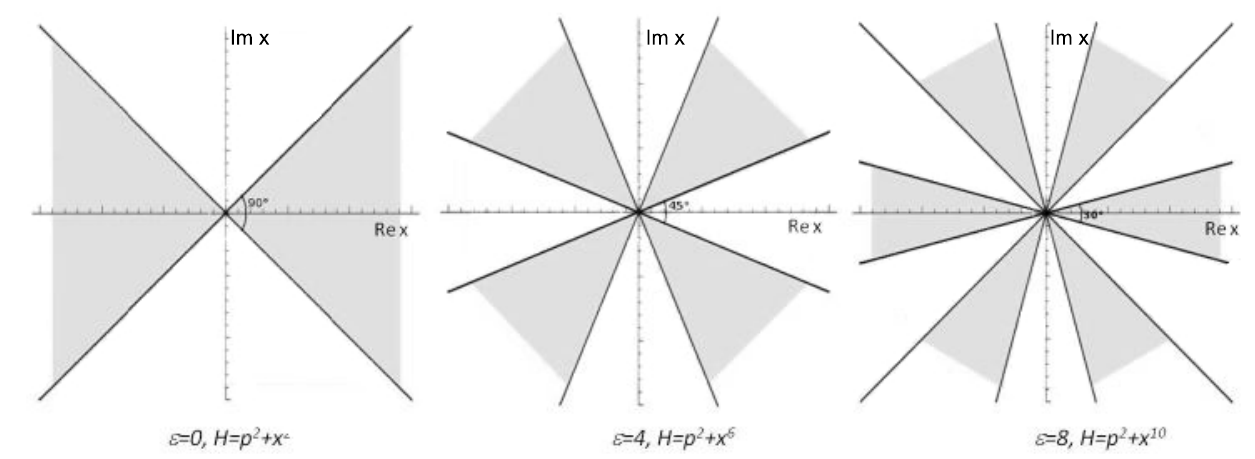}
			\end{center}
			\caption{Regions of convergence (Stokes' wedges) for even values of $\epsilon$ that are divisible by four, i.e. $\epsilon = 0,4,8$.}
			\label{fig:PicQM2} 
		\end{figure}		
		\end{widetext}

The case in which $\epsilon$ is even and divisible by four, i.e. $\epsilon=0,4,8,...$ leads to the Hamiltonians
$H_I=p^2+x^2; p^2+x^6$ and $p^2+x^{10}$ respectively. The Stokes' and anti-Stokes' lines can be easily determined and lead to the convergence patterns in the complex plane depicted in Fig.~\ref{fig:PicQM2}. In this figure, it is evident that a spectrum of real eigenvalues arises from conventional boundary conditions along the real axis, a feature expected from Hamiltonians that are Hermitian and bounded from below. This solution is the only one that can exist when $\epsilon=0$. However, for higher values of $\epsilon$ further real spectra can develop by considering boundary conditions that are  $\mathcal{PT}$-symmetric. For $\epsilon=4$ one additional pair of $\mathcal{PT}$-symmetric regions develops (from turning points that fall within the shaded wedges below the real axis), while for $\epsilon=8$, a pair of shaded wedges and a pair of unshaded wedges lie $\mathcal{PT}$-symmetrically to one another, leading to two further possible spectra of real eigenvalues. This result can be easily verified by evaluating the turning points in each case. For $\epsilon=4$ or $V(x)=x^6$, there are 6 possible turning points, which group into three pairs of $\mathcal{PT}$-symmetric values. Since the spectra resulting from the turning points in the upper half-plane (in the $\mathcal{PT}$-symmetric shaded wedges) are the same as those obtained from the turning points in the lower half-plane, only two distinct spectra are obtained. This reflection symmetry in the structure of the upper and lower half-planes is also evident for $\epsilon=8$, for which the potential $V(x)=x^{10}$ yields 5 pairs of $\mathcal{PT}$-symmetric turning points, lying in the $\mathcal{PT}$-symmetric Stokes' wedges shown in Fig.~\ref{fig:PicQM2}. Thus, in general, the number of distinct spectra containing real eigenvalues that may be associated with these Hamiltonians is $N=1+\epsilon/4$. The lowest spectrum can be calculated via Eq.~(\ref{eq:engamma}) to be $E_n\propto n+1/2$, $E_n\propto (n+1/2)^{3/2}$ and $E_n\propto (n+1/2)^{5/3}$, for $\epsilon=0,4$ and $8$ respectively. The ratios of the higher order spectra to the lower order ones are given as $R=2.83:1$ and $R=7.09:1.43:1$ for $\epsilon=4$ and $8$ respectively.

\begin{figure}[h]
			\begin{center}
			\includegraphics{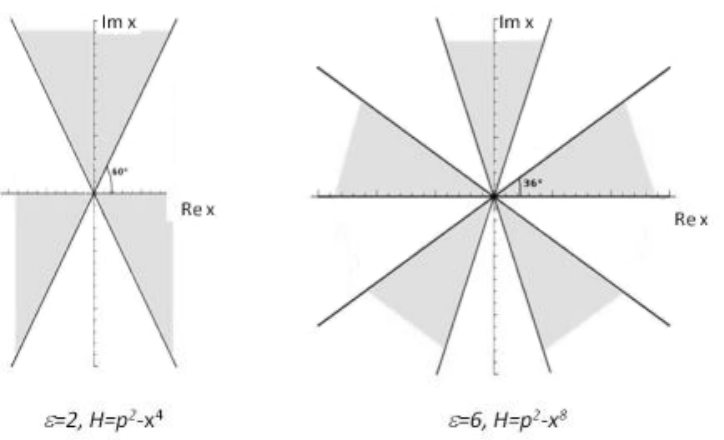}
			\end{center}
			\caption{Regions of convergence (Stokes' wedges) for even values of $\epsilon$ that are not divisible by four, for example $\epsilon = 2,6$.}
			\label{fig:PicQM3} 
		\end{figure}

Finally, we turn to the case where $\epsilon$ is even but not divisible by four, for example, $\epsilon=2$ or $6$. This choice leads to the Hamiltonians $H_I=p^2-x^4$ and $H_I=p^2-x^8$, which display the convergence patterns in the complex plane that are shown in Fig.~\ref{fig:PicQM3} and determined from Eqs.~(\ref{eq:psiHI}) to (\ref{eq:antistokes}). Although this class of Hamiltonians is Hermitian, it is unbounded below. This fact is reflected in the figure, as the real axis is a  Stokes' line on which the controlling factor of the asymptotic wavefunction is purely oscillatory: there can be no real (discrete) spectrum arising from boundary conditions given on this axis. Nonetheless, there {\it can} be spectra containing real eigenvalues associated with these Hamiltonians: for $\epsilon=2$, the shaded Stokes' wedges represent  $\mathcal{PT}$-symmetric regions in which the controlling factor of the asymptotic wavefunction converges to zero, and which can thus give rise to a spectrum of real eigenvalues. For $\epsilon=6$, there are two pairs of Stokes' wedges below the real axis, the shaded and the unshaded, which lie  $\mathcal{PT}$-symmetrically to one another and thus lead to two possible spectra of real eigenvalues. By symmetry arguments, results from the upper
 half-plane give rise to the same spectra as from the lower half-plane. These types of Hamiltonians thus have $N=(\epsilon+2)/4$ possible spectra of real eigenvalues. For the cases displayed in the figure, the discrete energy spectrum for the Hamiltonian $\epsilon=2$ is given as $E_n\propto (n+1/2)^{4/3}$, while for $\epsilon=6$, $E_n\propto (n+1/2)^{8/5}$ and the relation of the higher order solutions to the lower is $R=4.10:1$.

\subsection{Hamiltonians of the form $H_{II}=p^2+(x^2)^\delta$.}

\begin{figure}[h]
			\begin{center}
			\includegraphics{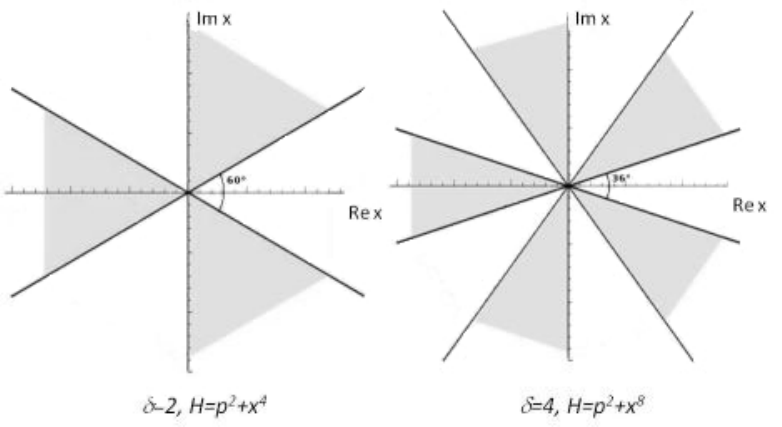}
			\end{center}
			\caption{Regions of convergence (Stokes' wedges) for $\delta = 2,4$, representing the Hamiltonians $H=p^2+x^4$ and $H=p^2+x^8$.}
			\label{fig:PicQM4}  
		\end{figure}

Hamiltonians discussed in the previous section fall formally into one single class, $H_I=p^2+x^2(ix)^\epsilon$. Viewing them individually for even values of $\epsilon$, one notes that the sign of the potential alternates, a feature that gives rise to a rich behaviour.  For real values of the potential, it is interesting to note how a change of sign alters the distribution of possible real (discrete) spectra. For this reason, we evaluate the spectra for Hamiltonians that are determined by $H_{II}=p^2+(x^2)^\delta$, with a view to examining those Hamiltonians that are not described by $H_{I}$ for any value of $\epsilon$.  Now the controlling factor in the asymptotic form of the wavefunction from Eq.~(\ref{eq:WKBWavefunction}) behaves as

		\begin{equation}
			\psi(x)\sim \exp \left(\pm \frac{1}{\delta+1}x^{\delta+1}\right)
		\label{eq:wavefunctionHII}
		\end{equation}
in the large $x$ limit. The Stokes' lines then follow on regarding $x$ as a complex variable and setting the real part of the wavefunction to zero. One finds
		\begin{equation}
			\varphi_S=\left(k+\frac{1}{2}\right)\frac{\pi}{\delta+1},
		\label{eq:stokesII}
		\end{equation}
where $k\in Z$. The anti-Stokes' lines occur at angles
		\begin{equation}
			\varphi_A=\frac{k}{\delta+1}\pi.
		\label{eq:antistokesII}
		\end{equation}
We illustrate the results for the cases $\delta=2$ and $4$ in Fig.~\ref{fig:PicQM4}, which differ markedly from that of the previous Hamiltonians of Subsection A: in both cases $\mathcal{PT}$-symmetric Stokes' wedges lie along the real axis. Each wedge, however, arises from a different convergent solution and is therefore differently shaded. The combination, as is well-known in quantum mechanics, leads to a spectrum of real eigenvalues. For $\delta=2$ and $H_{II}=p^2+x^4$, this is the only real spectrum that exists. For $\delta=4$, a second pair of $\mathcal{PT}$-symmetric wedges has developed in which the wave function converges asymptotically to zero, leading to a second possible spectrum of real eigenvalues for $H_{II}=p^2+x^8 $. In general, the number of families  found for even $\delta$ is $N=\delta/2$. The case $\delta$ odd, for example $\delta=1,3$, maps onto the cases already discussed in Section A, with $\epsilon=0,4$. The number of spectra of real eigenvalues obtained in this case is $N=(\delta+1)/4$.

\subsection{Hamiltonians of the form $H_{III}=p^2-(x^2)^\mu$.}

One distinct case in the type of Hamiltonians that can be constructed and which has not yet been dealt with fully is given by 	$H_{III}=p^2-(x^2)^\mu$. Here, the controlling behaviour of the asymptotic wavefunction is given by 
		\begin{equation}
			\psi(x)\sim \exp\left({\pm i\frac{1}{\mu+1} x^{\mu+1}}\right)
		\label{eq:psiIII}
		\end{equation}
for large $x$. Regarding $x$ as complex leads to the Stokes' lines occurring at angles
		\begin{equation}
			\varphi_S= \frac{k\pi}{\mu+1}
		\label{eq:stokesIII}
		\end{equation}
and the anti-Stokes' lines at
		\begin{equation}
			\varphi_A=\left(k-\frac{1}{2}\right) \frac {\pi}{\mu +1}
		\label{eq:antistokesIII}
		\end{equation}

\begin{figure}[h]
			\begin{center}
			\includegraphics{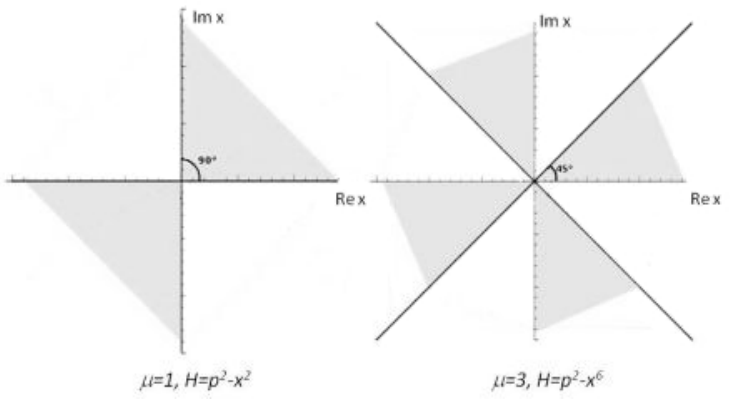}
			\end{center}
			\caption{Regions of convergence (Stokes' wedges) for $\mu = 1,3$, representing the Hamiltonians $H=p^2-x^2$ and $H=p^2-x^6$.}
			\label{fig:PicQM6} 
		\end{figure}

The  cases $\mu=1$ and $\mu=3$ are illustrated in Fig.~\ref{fig:PicQM6}. In  both of these cases, as with all the quantum-mechanical Hamiltonians, regions in which the controlling factor of the asymptotic wave function converges to zero exist in the entire complex $x$-plane. However for $\mu=1$, there are no noncontiguous 
$\mathcal{PT}$-symmetric wedges. In this case, no spectra with real eigenvalues exist. For the case $\mu=3$, one pair of noncontiguous 
$\mathcal{PT}$-symmetric wedges develops lying just below the real axis (left shaded, right unshaded). Here one spectrum with real eigenvalues can arise. In this case, the WKB approximation leads to $E_n\approx 2.81(n+1/2)^{3/2}$. For odd values of $\mu$, the number of real spectra that exist is given by $N=(\mu-1)/2$. Even values of $\mu$ correspond to the cases $\epsilon=2$ and $6$ respectively of $H_I$ detailed in Section A, leading to the result that the number of real discrete spectra in this case is  $N=\mu/2$.

	\section{Families of solutions in $\mathcal{PT}$-symmetric scalar field theory}
	\label{field}

		In this Section we review the analysis of $\mathcal{PT}$-symmetric scalar quantum field theory, following \cite{BK2010}. That is,  we show how to obtain  
		the regions of convergence of the partition function in the complex plane, explain how to deduce the number of real spectra from this image and finally obtain the 
		particle masses that are associated with each spectrum. In doing so, it is our aim to clarify the difference between the quantum-mechanical cases described in the previous section and the field-theoretical situation.

		We investigate systems with the set of Lagrangians given by
		\[
			\mathcal{L}=\frac{1}{2}(\partial\phi(x))^2+\frac{g}{4n+2}\phi^{4n+2}(x)-J(x)\phi(x)
		\]
		for different values of $n$. The partition function associated with this Lagrangian is given by 
		\[
			\begin{split}
				Z[J(x)]=\langle 0|0\rangle=&\int_C \mathcal{D}\phi(x)\exp\{-\int d^Dx'[\frac{1}{2}(\partial\phi(x'))^2 \\ &+\frac{g}{4n+2}\phi^{4n+2}(x')           -J(x')\phi(x')]\},
			\end{split}
		\]
		where $D$ is the dimensionality. The simplest case which is in direct analogy to the quantum-mechanical one is that in one dimension. The controlling behaviour of $Z[J(x)]$ is then determined by the integral 
		\begin{equation}
			\int_C d\varphi\exp\{-\varphi^{4n+2}\}.
		\label{eq:dphie}
		\end{equation}
		In a similar fashion to quantum mechanics, one can  analyse for which values of $\varphi$ the integral converges to zero when $|\varphi| \rightarrow \infty$.\\

\begin{widetext}

    		\begin{figure}[h]
			\begin{center}
			\includegraphics[scale=0.55, bb=0 0 685 282]{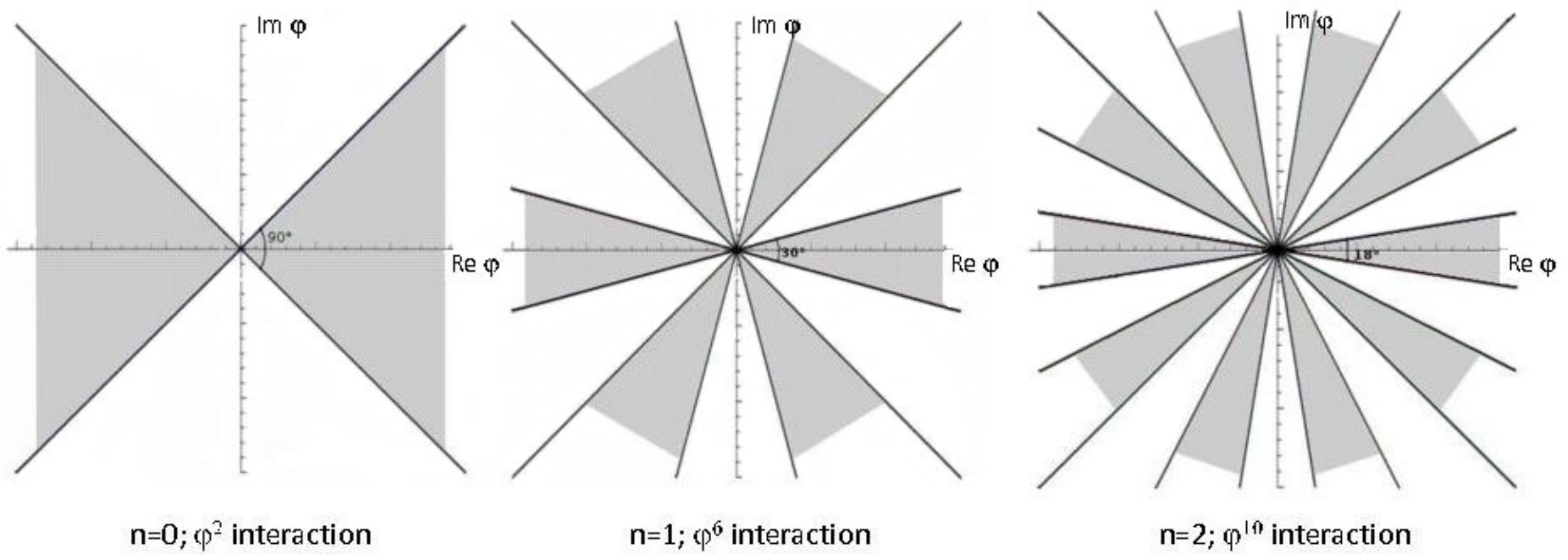}
			\end{center}
			\caption{Regions of convergence (Stokes' wedges) for $n = 0,1,2$}
			\label{fig:PicFT1} 
		\end{figure} 
\end{widetext}
		
Furthermore, it is possible to determine the energy spectra associated with $\mathcal{L}$ and thus the particle masses. To do so, we follow the treatment given in \cite{BK2010}. The field equation is deduced from the Lagrangian to be
		\[
			-\partial^2\phi(x)+g\phi^{4n+1}(x)=J(x).
		\]
		Taking the expectation value of this equation and using the defining relations for the vacuum expectation values and the connected Green functions,
		\[
			\langle\phi^n(x)\rangle=\frac{\delta^n Z[J(x)]}{\delta J^n(x)} \text{ and }
			G_n(x,x,...)=\frac{\delta^n \ln Z[J(x)]}{\delta J^n(x)},
		\]
		one can derive the infinite series of coupled Schwinger-Dyson equations. Truncating Green functions of third and higher order, one finally obtains two equations for $G_1$ and $G_2$, the latter of which determines  
		the mass $M$:
		\begin{equation}
                He_{2n+1}\left(iG_1/\sqrt{G_2(0)}\right)=0,
		\label{eq:g1}
		\end{equation} 
where $He_{2n+1}$ are Hermite polynomials of order $2n+1$, and
		\begin{equation}
                (-\nabla^2 + M^2)G_2(x-y) = \delta^D(x-y), 
	        \label{eq:g2mass}
		\end{equation} 
with
		\begin{equation}
                M^2=\left(G_2(0)\right)^{2n}He^\prime_{4n+1}\left(iG_1/\sqrt{G_2(0)}\right).
	        \label{eq:masssquared}
		\end{equation} 
 Solving Eqs.~(\ref{eq:g1}) - (\ref{eq:masssquared}) for $M$ and $G_1$, one generally obtains more than one real and positive solution for the mass $M$, while $G_1$ can be $0$,	 or negative and purely imaginary. This latter case represents the
 $\mathcal{PT}$-symmetric solution of the system.\\

For completeness, we consider first the case when $n$ takes on integral values, i.e. $n=1,2,3...$. The region in which Eq.~(\ref{eq:dphie}) converges is  again determined by the Stokes' or anti-Stokes' lines associated with this integral. Writing $\varphi=|\varphi|e^{i\vartheta}$, it follows that these angles occur at  \cite{BK2010} 
		\begin{equation}
		\vartheta_S=\left(\frac{k+1/2}{4n+2}\right)\pi
		\label{eq:theta_S}
		\end{equation}
and
		\begin{equation}
		\vartheta_A=\frac {k}{4n+2} \pi,
		\label{eq:theta_A}
		\end{equation} 
where $k$ is an integer. The controlling parts of the field-theoretical integrals - unlike their quantum-mechanical counterparts - do not converge to zero in {\it all} regions of the complex $\varphi$ plane. The convergence patterns are shown in Fig.~\ref{fig:PicFT1} for the cases $n=0, 1, 2$. Each case has a convergent region along the real axis, yielding one real energy spectrum, or, for the ground state, the particle mass. In addition, however, the case $n=1$ develops further regions of convergence, which lie $\mathcal{PT}$-symmetrically to one another and thus generate a further spectrum leading to a further mass that has the ratio $R=1.62:1$ to the first mass. When $n=2$, a second extra real energy spectrum develops, so that there are three in total, leading to  masses lying in the ratio $R=2.26:1.19:1$ to one another. The region of convergence for $n=0$ strongly resembles the quantum-mechanical case depicted in Fig.~\ref{fig:PicQM2} when $\epsilon=0$. The fact that there is no convergent solution in the field-theoretic case in the vertical wedges does not play any role in determining the number of real spectra. These two cases are equivalent. For the case that $n=1$, however, the distribution of Stokes' wedges shown in Fig.~\ref{fig:PicFT1} resembles the case $\epsilon=8$ in Fig.~\ref{fig:PicQM2}, but has one less available pair of $\mathcal{PT}$-symmetric wedges to construct a real spectrum. The quantum-mechanical case thus admits one more real spectrum than the field-theoretical case with a similar wedge structure. The number of real spectra in the field-theoretical case is given by $N=n+1$.

\begin{figure}[h!]
			\begin{center}
			\includegraphics[scale=0.55, bb=0 0 685 282]{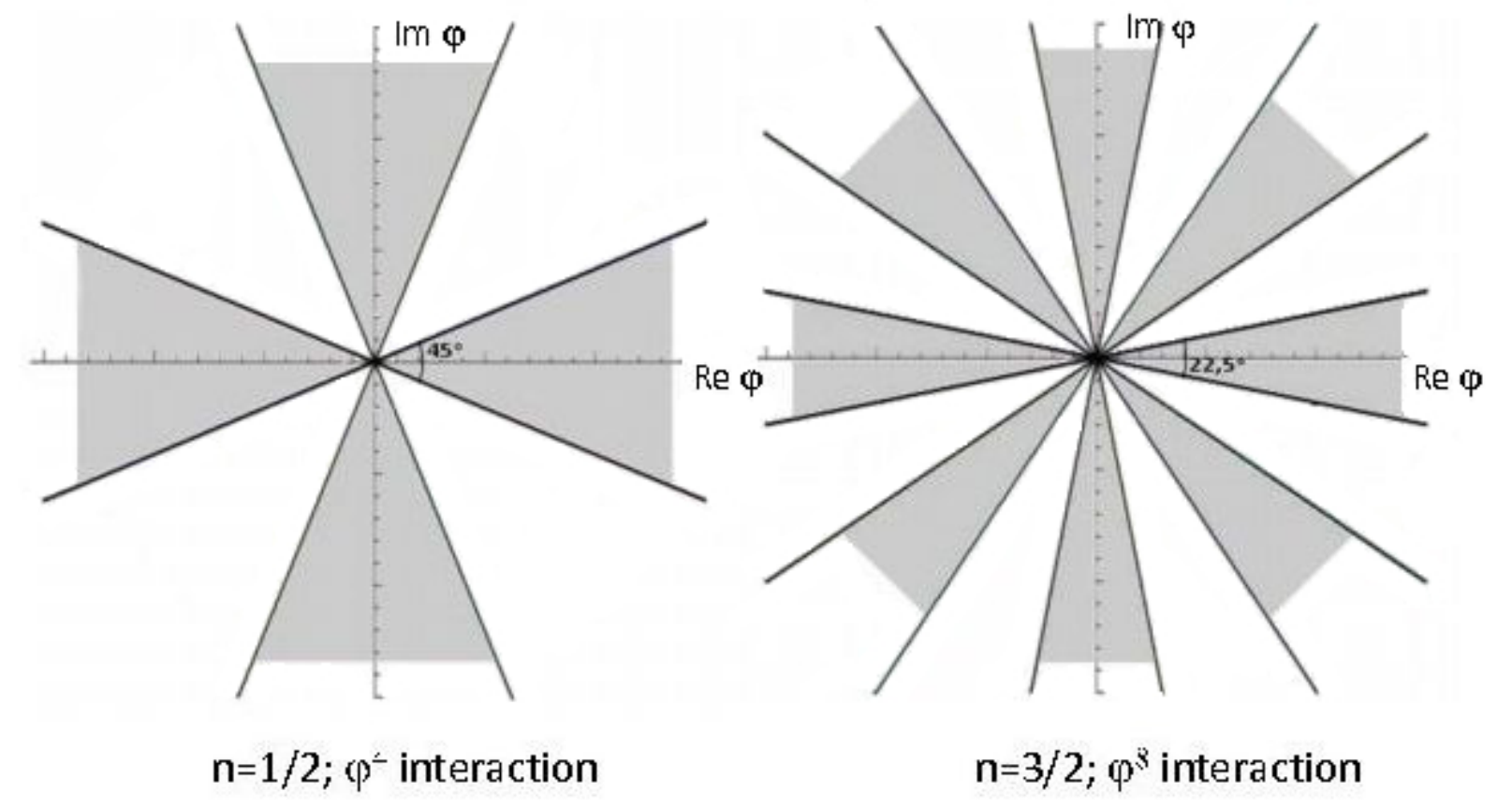}
			\end{center}
			\caption{Regions of convergence (Stokes' wedges) for $n = 1/2$ and $3/2$.}
			\label{fig:PicFT2}  
		\end{figure}

For half-integer values of $n$, such as $n=1/2$ or $n=3/2$, the picture shown in Fig.~\ref{fig:PicFT2} emerges. For $n=1/2$, only one real energy spectrum can exist, while for $n=3/2$ an additional $\mathcal{PT}$-symmetrical spectrum can be found, with mass ratio $R=1.31:1$ between the two solutions. In general, for half-integer values of $n$ there are $N=n+1/2$ possible real spectra.

\begin{figure}[h!]
			\begin{center}
			\includegraphics[scale=0.55, bb=0 0 685 282]{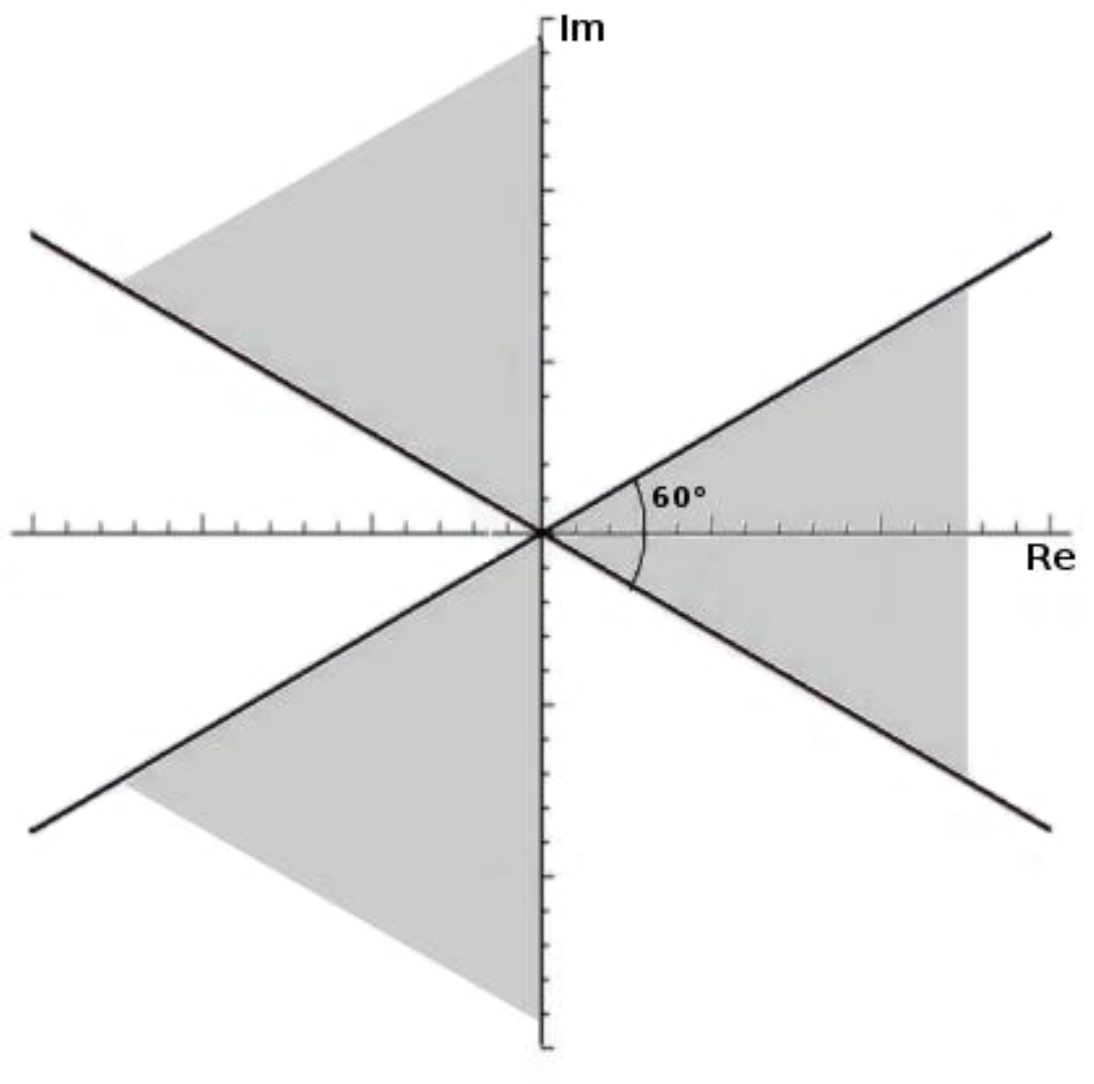}
			\end{center}
			\caption{Regions of convergence (Stokes' wedges) for $n = 1/4$.}
			\label{fig:fi3}   
		\end{figure}

Finally, we note that there are no $\mathcal{PT}$-symmetric solutions for values of $n=1/4, 3/4, 5/4, \dots$ corresponding to interactions of the form $\varphi^3$, $\varphi^5$ and $\varphi^7$ respectively. This can be seen immediately on examining the pattern of Stokes' wedges for example, for $n=1/4$, shown in Fig.~\ref{fig:fi3}. 

\section{Summary and conclusions}
	\label{summary}

In this paper, we have described the classification of families of (positive) real spectra associated with the quantum-mechanical Hamiltonians  $H_I=p^2+x^2(ix)^\epsilon$, 			$H_{II}=p^2+(x^2)^\delta$ and $H_{III}=p^2-(x^2)^\mu$, with $\epsilon, \delta$ and $\mu >0; \epsilon, \delta, \mu \in Z$ within the WKB approximation and for a scalar field theory  $\mathcal{L}=\frac{1}{2}(\partial\phi(x))^2+g/(4n+2)\phi^{4n+2}(x)-J(x)\phi(x)$ within the functional integral formalism. In the quantum-mechanical case, we have shown that the number of real spectra depends on the number of pairs of $\mathcal{PT}$-symmetric turning points that can be found in the complex plane. This in turn depends on the type of Hamiltonian and the form of the interaction potential. The classification of the number of $\mathcal{PT}$-symmetric turning points can be reinterpreted in terms of the position and number of Stokes' wedges in the complex plane, leading to the rule that each pair of noncontiguous $\mathcal{PT}$-symmetric Stokes' wedges may result in a positive energy spectrum containing real eigenvalues; two pairs of Stokes' wedges which are symmetric to the real axis may however only be counted once. Thus, by simply examining the Stokes' wedges in the complex plane, one can deduce the number of  spectra of real eigenvalues that can occur. One recovers the usual result that Dirac-Hermitian Hamiltonians have a spectrum of real eigenvalues due to "normal" boundary conditions placed along the real axis. However, these Hamiltonians can give rise to additional real spectra if additional $\mathcal{PT}$-symmetrical regions of convergence exist in the complex plane. Non-Hermitian, but $\mathcal{PT}$-symmetric Hamiltonians can also possess a spectrum of real eigenvalues (in fact several), while the class of Dirac-Hermitian Hamiltonians that appear unbounded from below can also produce several such spectra or families. This is because the Hamiltonians must always be solved together with a set of given boundary conditions.   For all cases, the spectra can be evaluated simply within the WKB approximation and the ratios of the different spectra can be found.  

A similar analysis of the scalar field theory using functional integrals shows that again the structure of the Stokes' wedges in the complex plane suffices to give information on the number of possible real spectra. However, fewer real spectra are available in cases which appear formally similar to their quantum-mechanical counterparts, as the regions of convergence of the asymptotic controlling factor in the complex plane are restricted by the sign of the power of the action, as it occurs in the generating functional. The ratio of the masses (ground state energies) of the individual spectra can also be simply evaluated. Here the approximation lies in the truncation of the Dyson-Schwinger equations.

Finally, it is appropriate to stress the limitations of the analysis given here. In the quantum-mechanical cases, it is unclear how well the WKB approximation reproduces the actual energy eigenvalues, which can only be calculated definitively numerically. Usually the result is excellent for the lowest states, see \cite{B2007}. It is, however, not possible to use it to describe the spontaneous breakdown of $\mathcal{PT}$-symmetry, and the associated phase transition. Nor can it be used to address deeper questions such as whether a particular spectrum has a finite or infinite number of real eigenvalues. More powerful techniques are needed. Some useful references that go beyond the WKB approach can be found for example in \cite{DDT2001} and \cite{DMT2005}.

\section{Acknowledgments}
This work has been performed at the Heidelberg Graduate School of Fundamental Physics, Germany, to which we are grateful. One of us, SPK, thanks Carl Bender for stimulating discussions. One of us, SPK, also thanks the Physics Department of the University of Johannesburg for its kind hospitality while preparing this manuscript.

\end{document}